\documentclass[onecolumn,preprint]{revtex4-1}
\usepackage{amssymb}
\usepackage{amsfonts}
\usepackage{amsmath}
\usepackage{graphicx}
\usepackage{epstopdf}
\usepackage{float}

\begin{document}

\preprint{HEP/123-qed}
\title{Scalar bosons under the influence of noninertial effects in the
cosmic string spacetime}
\author{L. C. N. Santos}
\author{C. C. Barros Jr.}
\affiliation{Depto de F\'{\i}sica - CFM - Universidade Federal de Santa Catarina, CP. 476
- CEP 88.040 - 900, Florian\'{o}polis - SC - Brazil}

\begin{abstract}
In this paper we present two different classes of solutions for the
Klein-Gordon equation in the presence of a scalar potential under the
influence of noninertial effects in the cosmic string spacetime. We show
that noninertial effects restrict the physical region of the spacetime where
the particle can be placed,\ and furthermore that the energy levels are
shifted by these effects. In addition, we show that the presence of a
Coulomb-like scalar potential allows the formation of bound states when the
Klein-Gordon equation is considered in this kind of spacetime.
\end{abstract}

\volumeyear{}
\volumenumber{}
\issuenumber{}
\eid{identifier}
\startpage{1}
\endpage{10}
\maketitle

\section{\protect\bigskip \protect\bigskip introduction}

In the past few years, the scientific interest in the study of gravitational
effects on quantum mechanical systems has been renewed and many systems has
been studied \cite%
{Castro1,Castro2,parker1,parker2,parker3,parker4,parker5,hcurvo1,Barros1,chandrasekharansatz,cohen,neutrino1}%
. For example, in \cite{parker1} it was shown that the energy spectrum
associated with one-electron atoms in an arbitrary curved spacetime is
different from the one obtained in a usual flat Minkowski spacetime. The
energy levels are shifted by the gravitational field and the effects of the
curvature appear as perturbations in the relativistic fine structure. In 
\cite{neutrino1} neutrinos have been studied in this context, and in \cite%
{santos1}, the effects of magnetic fields in the metric have been considered.

Another kind of system that may be investigated with this purpose are the
cosmic strings. Cosmic strings are very interesting systems, that are
supposed to be formed during a symmetry breaking phase transition in the
early universe \cite{string1,string2,string3,string4,string5,string6,string7}
and may be considered as topological defects in the spacetime structure. The
spacetime around them is locally flat, but this is not a global propriety.
The presence of such kind of topological defects can also influence the
behavior of a quantum mechanical system, as in \cite{hcurvo1}, where the
Dirac equation has been solved in the presence of a Coulomb and a scalar
potentials in the cosmic string spacetime and it was shown that its presence
destroys the degeneracy of all the energy le-vels. In \cite{string8},
solutions of the Klein-Gordon equation in G\"{o}del-type spaces with an
embedded cosmic string are considered, and it was shown that the presence of
topological defects in the spacetime breaks the degeneracy of energy levels
in the cases of Som--Raychaudhuri, spherical and hyperbolic G\"{o}del
solutions.

Noninertial effects of rotating frames on physical systems is another kind
of aspect that have been studied in many works in the literature, as for
example in \cite{bakke2,bakke3}. For instance, the Mashhoon effect is the
coupling of the spin of the particles with the angular velocity of the
rotating frames and it arises from the influence of these noninertial frames
when interference effects are considered \cite{inercial4}. In \cite%
{inercial8}, the Landau quantization for neutral particles with a permanent
magnetic dipole moment has been studied, and it has been shown that the
noninertial effects modify the cyclotron frequency. The molecular
Aharonov-Carmi (A-C) effect has been considered for $C_{60}$ molecules in 
\cite{inercial9}, and the energy shift of the valence electrons due to the
molecular rotation has been calculated.

On the other hand, the study of scalar potentials in quantum mechanics is
very important in order to establish basic proprieties of several systems of
interest. The absence of Klein's paradox\ for scalar potentials is an
important feature of this approach. In this way, if a potential is
vectorlike, the possibility of a tunneling solution arises, but this
situation does not occur when the confining potential is a scalar \cite%
{escalar1}.

In addition, the scalar potential can simulate effective masses, as it is
known from the nuclear or the solid state physics, and the effect of the
spatial dependence of these effective masses has been found in many physical
systems. Recently, the study of some of these systems has attracted the
attention from many authors \cite%
{massa1,massa2,massa3,massa4,massa5,massa7,massa8,massa9,massa10}.
Applications have been found in fields such as the electronic properties of
the semiconductors \cite{massa12}, $^{3}$He clusters \cite{massa13}, quantum
liquids \cite{massa14}, semiconductor heterostructures \cite{massa15} and
others. One side of this research is concerned with the development of
methods and techniques for studying how do these mass variations affect the
dynamics of quantum systems.

Solutions of nonrelativistic wave equations with spatial dependence of the
effective mass have been obtained for many systems. The Schr\"{o}dinger
equation with a smooth mass and a step potential has been solved by Dekar et
al. \cite{massa16}, where the behavior of the transmission coefficient have
been compared to an abrupt step potential. Besides, the supersymmetric
quantum-mechanical formalism to the Schr\"{o}dinger equation has been
extended to describe particles characterized by position-dependent effective
masses \cite{massa17}. Exact solutions of a spatially dependent mass Dirac
equation via Laplace transformation method \cite{massa18} have been also
obtained \ explicitly. This method is an integral transform and recently
have been used to solve the Schr\"{o}dinger equation in non-relativistic
problems.

In the present paper, we will study spin-0 bosons in a cosmic string
spacetime by conside-ring the Klein-Gordon equation in the presence of a
Coulomb-like scalar potential $s\left( r\right) =\eta/r$, where $r=\sqrt{%
x^{2}+y^{2}}$ is the radial coordinate and $\eta$ a constant. In addition, a
rotating frame in the cosmic string spacetime will be considered, and \ we
will show that noninertial effects restrict the physical region of the
spacetime where the particle can be placed,\ and furthermore the energy
levels are shifted by the noninertial effects on the particle. This
interesting feature is an indicator of a deeper phenomenon: the coupling
between the angular quantum number and the angular velocity of the rotating
frame. In what follows, we show that the presence of a Coulomb-like scalar
potential can form bound states for the Klein-Gordon equation in this
spacetime.

The paper is organized as follows: In Section II, we will describe the
cosmic string spacetime and the transformation from spacetime coordinates to
rotating coordinates. In Section III, we will present a equation for spin-0
bosons in a cosmic string spacetime for the potential $V\left( r\right)
=\eta /r$ and in Section IV, we will obtain the numerical computation of the
Klein-Gordon by a root-finding procedure. Finally, Section V presents our
conclusions. In this work, we use natural units, $\hbar =c=G=1$.

\section{The cosmic string spacetime}

The metric of a cosmic string \ is a solution of Einstein's equations and it
describes a spacetime determined by an infinitely long straight string with
nonvanishing thickness. The string spacetime is assumed to be static and
cylindrically symmetric, and then, the metric re\-pre\-sen\-ting this system
is given by \cite{string5,Castro2}%
\begin{equation}
ds^{2}=-dt^{\prime 2}+dr^{\prime 2}+\alpha ^{2}r^{\prime 2}d\phi ^{\prime
2}+dz^{\prime 2},  \label{eq1}
\end{equation}%
where $\alpha =1-4G\mu $ and $\mu $ is the mass density of the string. In
this metric, the azimuthal angle range is $\phi ^{\prime }\in \lbrack 0,2\pi
)$ while the $r$ coordinate range is $r$ $\in \lbrack 0,\infty )$. The
parameter $\alpha $ may assume values in which $\alpha \leq 1$ or $\alpha >1$%
, and in this case, it corresponds to a spacetime with negative curvature.
Moreover $\alpha $ represents the deficit angle of the conical spacetime and 
$\alpha =1$ corresponds to a flat spacetime. In this work, we are interested
in studying the case $0<\alpha <1$.The generalization of this formulation
for a noninertial reference frame may be made by considering a coordinate
transformation%
\begin{equation}
t^{\prime }=t,\text{ \ \ }r^{\prime }=r,\text{ \ \ \ }\phi ^{\prime }=\phi
+\omega t,\text{ \ \ }z^{\prime }=z,  \label{eq2}
\end{equation}%
where $\omega $ is angular velocity of the rotating frame. In\-ser\-ting
this transformation into eq. (\ref{eq1}) we obtain the line element%
\begin{align}
ds^{2}& =-dt^{2}+dr^{2}+\alpha ^{2}r^{2}\left( d\phi +\omega dt\right)
^{2}+dz^{2}  \notag \\
& =-\left( 1-\alpha ^{2}r^{2}\omega ^{2}\right) dt^{2}+2\alpha
^{2}r^{2}\omega dtd\phi  \notag \\
& +dr^{2}+\alpha ^{2}r^{2}d\phi ^{2}+dz^{2},  \label{eq3}
\end{align}%
that may be associated with the covariant metric tensor 
\begin{equation*}
g_{\mu \eta }=\left( 
\begin{array}{cccc}
-\left( 1-\alpha ^{2}r^{2}\omega ^{2}\right) & 0 & 0 & \alpha ^{2}r^{2}\omega
\\ 
0 & 1 & 0 & 0 \\ 
0 & 0 & 1 & 0 \\ 
\alpha ^{2}r^{2}\omega & 0 & 0 & \alpha ^{2}r^{2}%
\end{array}%
\right) ,
\end{equation*}%
that is a non-diagonal metric tensor where the effects of the topology and
the rotation of the reference frame are taken into account. An interesting
feature of the equation (\ref{eq3}) is the condition 
\begin{equation}
0<r<1/\alpha \omega .  \label{eq4}
\end{equation}%
This condition is related to the fact that for $r>$ $1/\alpha \omega $ the
velocity of the particle is greater than the velocity of the light, for this
reason, it is convenient to restrict $r$ to the range (0,$1/\alpha \omega $%
). Thus, the wave function of the quantum particle must vanish at $%
r=1/\alpha \omega $ \ and this system presents two different classes of
solutions that depend on the value of the product $\alpha \omega $. The
first case is obtained by adopting the limit $\alpha \omega \ll 1$ ($%
1/\alpha \omega \rightarrow \infty $), \ and as a second case, we have
considered an arbitrary relation $\alpha \omega .$

\section{Klein-Gordon equation in the cosmic string spacetime}

In a flat Minkowski spacetime, the spin-0 particles are represented by the
usual Klein-Gordon equation. In this section we will present a equation for
these particles in a curved spacetime. In order to determine this equation
one may replace the ordinary derivatives by covariant derivatives in the
equation and the result is 
\begin{equation}
-\frac{1}{\sqrt{-g}}\partial _{\mu }\left( g^{\mu \nu }\sqrt{-g}\partial
_{\nu }\psi \right) +m^{2}\psi =0,  \label{eq5}
\end{equation}%
that is the Klein-Gordon equation in a curved spacetime \cite{birrel}, where 
$m$ is the particle mass and $g$ is the determinant of the metric tensor. An
arbitrary scalar potential may be taken into account by making a
modification in the mass term as $m\rightarrow m+V\left( r\right) $.
Substituting the mass term into (\ref{eq5}) we obtain the following
differential equation%
\begin{equation}
-\frac{1}{\sqrt{-g}}\partial _{\mu }\left( g^{\mu \nu }\sqrt{-g}\partial
_{\nu }\psi \right) +\left( m+V\right) ^{2}\psi =0.  \label{eq5b}
\end{equation}%
Solutions of equation (\ref{eq5b}) are, in general, very difficult to find
and the known ones are limited to a small set of potentials and spacetimes.
In the following, we will obtain two classes of solutions of the
Klein-Gordon equation in a cosmic string space for Coulomb-like scalar
potentials as an example. By considering the line element (\ref{eq3}), we
obtain%
\begin{equation*}
\left[ -\partial _{t}^{2}+\partial _{r}^{2}+\frac{1}{r}\partial
_{r}+\partial _{z}^{2}+2\omega \partial _{t}\partial _{\phi }\right.
\end{equation*}%
\begin{equation}
\left. +\left( \frac{1}{\alpha ^{2}r^{2}}-\omega ^{2}\right) \partial _{\phi
}^{2}-\left( m+V\right) ^{2}\right] \psi =0,  \label{eq6}
\end{equation}%
that is the Klein-Gordon equation in the cosmic string spacetime. The
equation (\ref{eq6}) is independent of $t$, $z$ and $\phi $, so it is
reasonable to write the solution as%
\begin{equation}
\psi \left( t,r,z,\phi \right) =e^{-i\varepsilon t}e^{il\phi
}e^{ip_{z}z}R\left( r\right) ,  \label{eq7}
\end{equation}%
where $l=0,\pm 1,\pm 2,\pm 3$, $\varepsilon $ can be interpreted as the
energy of the particle and $p_{z}$ is the momentum in the $z$ direction.
Substituting (\ref{eq7}) into Eq. (\ref{eq6}), we obtain the radial
differential equation%
\begin{equation*}
\left[ \frac{d^{2}}{dr^{2}}+\frac{1}{r}\frac{d}{dr}-\frac{l^{2}}{\alpha
^{2}r^{2}}\right. -\left( m+V\right) ^{2}
\end{equation*}%
\begin{equation}
\left. +\left( \varepsilon +\omega l\right) ^{2}-p_{z}^{2}\right] R\left(
r\right) =0.  \label{eq8}
\end{equation}
From equation $\left( \ref{eq8}\right) $, we can see that the term $\omega l$
works as a time--like vector potential, i.e., the noninertial effect of
rotating frames in the Klein--Gordon equation is equivalent to a time--like
vector potential.

It is possible to define a four-current $J_{\mu }$ in terms of our preceding
results. The equation (\ref{eq5b}) \ can be expressed as%
\begin{equation}
-\nabla ^{\nu }\nabla _{\nu }\psi +\left( m+V\right) ^{2}\psi =0,
\label{eq5c}
\end{equation}%
where $\nabla _{\nu }$ is the covariant derivative.

The conservation law for $J_{\mu }$ follows from the procedure of
multiplying $\left( \ref{eq5c}\right) $ and its complex conjugate by $\psi $
from the left and by $\psi ^{\ast }$ from the right, respectively. The sum
of those resulting equations leads to%
\begin{equation}
\nabla ^{\nu }J_{\nu }=0,  \label{eq5d}
\end{equation}%
where 
\begin{equation}
J_{\nu }=\frac{i}{2m}\left( \psi ^{\ast }\nabla _{\nu }\psi -\psi \nabla
_{\nu }\psi ^{\ast }\right) .  \label{eq5e}
\end{equation}%
Hence, we can identify $J^{0}$ as a charge density and, thus, we have
particles with charge $+e$ and particles with charge $-e$ in accordance with
the existence of particles and antiparticles in the theory. From (\ref{eq7})
and (\ref{eq5e}), we can see that the charge density (covariant component)
is given by%
\begin{equation}
J_{0}=\frac{\varepsilon }{m}\left\vert \psi \right\vert ^{2}.  \label{eq5f}
\end{equation}

\bigskip

\subsection{ potential $V\left( r\right) =\protect\eta/r$.}

We want to solve the radial equation for a scalar potential of the type $%
V\left( r\right) =\eta /r$, where $\eta $ is a constant. So, substituting
this potential into Eq. (\ref{eq8}) we obtain the following expression \ 
\begin{equation}
\left[ \frac{d^{2}}{dr^{2}}+\frac{1}{r}\frac{d}{dr}-\frac{\beta ^{2}}{r^{2}}-%
\frac{2\gamma }{r}-\delta ^{2}\right] R\left( r\right) =0,  \label{eq9}
\end{equation}%
where 
\begin{equation}
\delta ^{2}=m^{2}+p_{z}^{2}-\left( \varepsilon +\omega l\right) ^{2},\text{
\ }\beta ^{2}=l^{2}/\alpha ^{2}+\eta ^{2},\text{ \ }\gamma =m\eta .
\label{eq9b}
\end{equation}%
In order to investigate the solutions of the Eq. (\ref{eq9}), we will
consider a transformation of the radial coordinate 
\begin{equation}
\rho =2\delta r,  \label{eq10}
\end{equation}%
and as a result, the equation will take the form%
\begin{equation}
\left[ \frac{d^{2}}{d\rho ^{2}}+\frac{1}{\rho }\frac{dR}{d\rho }-\frac{\beta
^{2}}{\rho ^{2}}-\frac{\gamma }{\delta \rho }-\frac{1}{4}\right] R\left(
\rho \right) =0.  \label{eq11}
\end{equation}%
Normalizable eigenfunctions may be obtained if we propose the solution%
\begin{equation}
R\left( \rho \right) =\rho ^{\beta }e^{-\frac{\rho }{2}}F\left( \rho \right)
,  \label{eq12}
\end{equation}%
and substituting (\ref{eq12}) into eq. (\ref{eq11}) and remembering that the
parameter $\beta $ is a constant, we obtain the differential equation
associated with the radial solution%
\begin{equation}
\rho \frac{d^{2}F}{d\rho ^{2}}+\left( 2\beta +1-\rho \right) \frac{dF}{d\rho 
}-\left( \beta +\frac{\gamma }{\delta }+\frac{1}{2}\right) F=0.  \label{eq13}
\end{equation}%
This is the confluent hypergeometric equation, which is a second order
linear \ homogeneous differential equation. The general solution of Eq. (\ref%
{eq13}) is given by 
\begin{equation}
F=AM\left( a,b;\rho \right) +B\rho ^{1-b}M\left( a-b+1,2-b;\rho \right)
\label{eq13b}
\end{equation}%
where 
\begin{eqnarray*}
b &=&2\beta +1, \\
a &=&\beta +\frac{\gamma }{\delta }+\frac{1}{2},
\end{eqnarray*}%
with arbitrary constants $A$ and $B$. Here, the confluent hypergeometric
function $M\left( a,b;\rho \right) $ is denoted by 
\begin{equation}
M\left( a,b;\rho \right) =_{1}F_{1}\left( a,b;\rho \right) .  \label{eq14}
\end{equation}%
We can see that the second term in $\left( \ref{eq13b}\right) $ has a
singular point at $\rho =0$, so we set $B=0$.

\subsection{Limit $\protect\alpha\protect\omega\ll1$ ($1/\protect\alpha%
\protect\omega\rightarrow\infty$)}

\begin{figure}[b]
\includegraphics[scale=0.7]{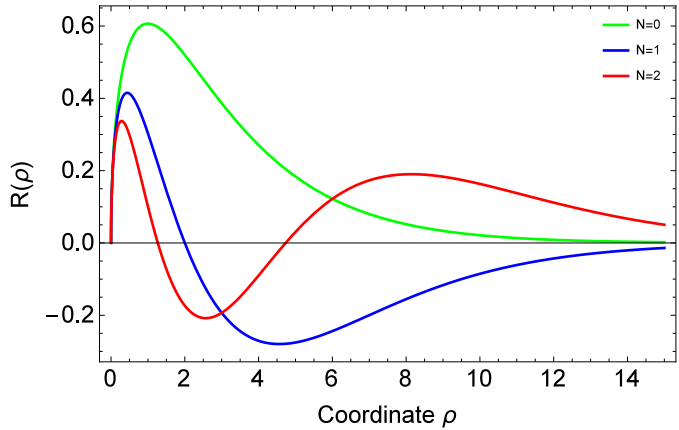}\newline
\caption{\label{fig1}The plots of radial coordinate $R$\ as a
function of variable $\rho$\ displayed for three different values of $N$ with
the parameters $\alpha=0.8$, $\omega=0.1$, $\eta=-0.5$ and $l=0.$}
\end{figure}

Considering now the limit $\alpha \omega \ll 1$, that is a slow rotation
regime, the boundary condition implies that the expression%
\begin{equation}
_{1}F_{1}\left( \beta +\frac{\gamma }{\delta }+\frac{1}{2},2\beta +1;\frac{%
2\delta }{\alpha \omega }\rightarrow \infty \right)  \label{eq15}
\end{equation}%
must be finite. Due to asymptotic behavior of solution, it is necessary that
the hypergeometric function be a polynomial function of degree $N$, that
means that the parameter $\beta +\frac{\gamma }{\delta }+\frac{1}{2}$ should
be a negative integer. This condition implies that 
\begin{equation}
\beta +\frac{\gamma }{\delta }+\frac{1}{2}=-N,\text{ \ \ \ }N=0,1,2,...\text{
,}  \label{eq16}
\end{equation}%
and combining this equation and equation $\left( \ref{eq9b}\right) $ we
finally obtain the energy spectrum%
\begin{equation}
\varepsilon _{\pm }=-\omega \left\vert l\right\vert \pm \sqrt{p_{z}^{2}+m^{2}%
\left[ 1-\frac{\eta ^{2}}{\left( N+\frac{1}{2}+\sqrt{l^{2}/\alpha ^{2}+\eta
^{2}}\right) ^{2}}\right] .}  \label{eq17}
\end{equation}

\begin{figure}[h]
\includegraphics[scale=0.7]{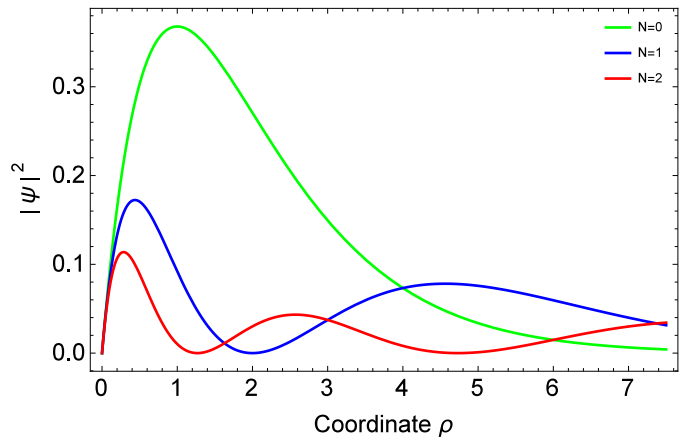}\newline
\caption{The plots of $\left\vert \psi\right\vert ^{2}$\ \ as functions of $\rho$ 
displayed for three different values of $N$ with
the parameters $\alpha=0.8$, $\omega=0.1$, $\eta=-0.5$ and $l=0$.}
\end{figure}

This is the energy spectrum for both particle ($\varepsilon _{+}$) and
antiparticle ($\varepsilon _{-}$).We can see that the energy spectrum
associated with the Klein-Gordon equation in cosmic string space for the
Coulomb-like scalar potential depends on $\alpha $, the deficit angle of the
conical space-time. For $l=0$ and $\omega =0$ the discrete set of energies
are symmetrical about $\varepsilon =0$. In this way, the presence of
noninertial effects of rotating frames in spacetime breaks the symmetry of
energy levels about $\varepsilon =0$.

\begin{figure}[h]
\includegraphics[scale=0.4]{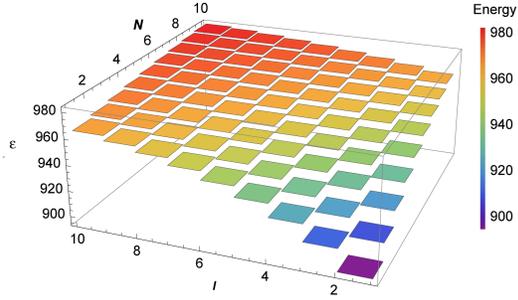}\newline
\caption{The plots of particle energy spectrum $\varepsilon$ as the function
of variables $N$ and $l$ for $\alpha=0.9$, $\omega=0.01$, $\eta=5$, and $m=1000$. }
\end{figure}

As it was said before, the parameter $\omega$ is the angular velocity of the
rotating frame, so we can observe that the energy levels are shifted by
noninertial effects on the particle.

\begin{figure}[h]
\includegraphics[scale=0.4]{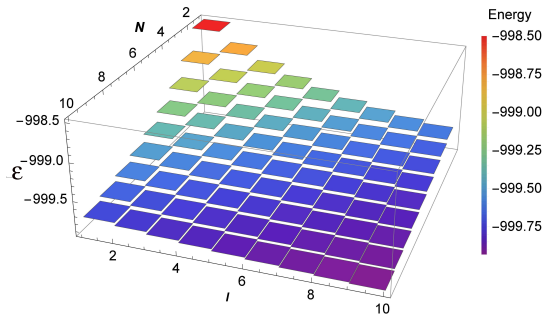}\newline
\caption{The plots of negative energy spectrum $\varepsilon$ as the function
of variables $N$ and $l$ for $\alpha=0.9$, $\omega=0.01$, $\eta=0.5$, and $m=1000$.}
\end{figure}

As it may be seen in Fig. 1, the radial solution $R\left( \rho \right) $
decreases with the coordinate $\rho $ and becomes negligible far away from
the cosmic string as $\rho \rightarrow \infty $. In Fig.2, we illustrate the
plots of $\left\vert \psi \right\vert ^{2}$ as a function of variable $\rho $
for three different values of $N$. The plots of the energy spectrum $%
\varepsilon $ as function of the variables $N$ and $l$ are shown in Figures
3 and 4.

\begin{figure}[H]
\includegraphics[scale=1]{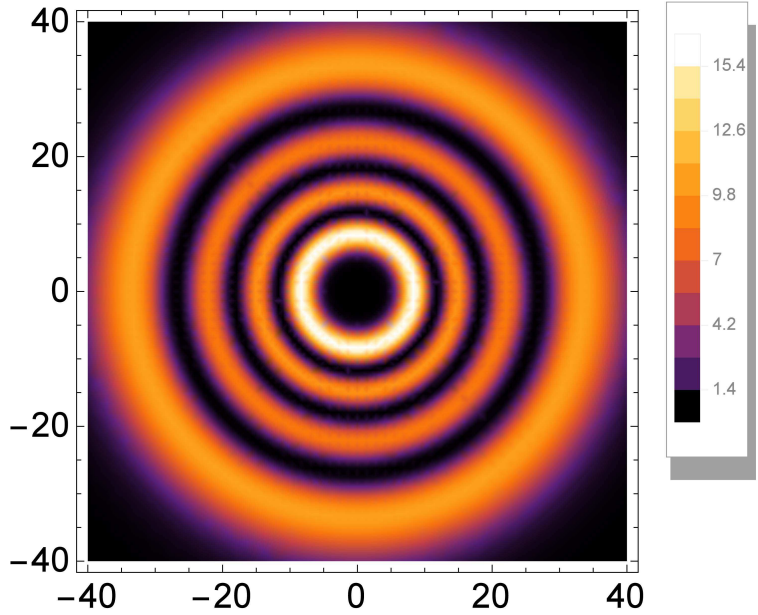}\newline
\caption{Plot of $\ \left\vert \psi\right\vert ^{2}$  in polar coordinates where $\rho=\sqrt{x^{2}+y^{2}}$, $x$  is the horizontal axis and  $y$ is the vertical exis for $N=3$ ,
$m=100$, $\alpha=0.5$, $\omega=0.1$, and $\eta=-0.4$.}
\end{figure}

The Behavior of $\ \left\vert \psi \right\vert ^{2}$ for $N=3$ , $m=100$,
and $\alpha =0.5$ in polar coordinates, is plotted in Figure 5, and we
can see that the scalar bosons tend to be better localized at the white
region.

\section{\protect\bigskip Arbitrary $\protect\omega \protect\alpha $}

Now if we do not impose condition $\alpha \omega \ll 1$ for the scalar
particle equation, as it was pointed before, due to noninertial effects, the
physical condition implies that the eigenfunction vanishes at $r\rightarrow
1/\alpha \omega $, that means 
\begin{equation}
_{1}F_{1}\left( \beta +\frac{\gamma }{\delta }+\frac{1}{2},2\beta +1;\frac{%
2\delta }{\alpha \omega }\right) =0.  \label{eq18}
\end{equation}%
The quantization condition in $\left( \ref{eq18}\right) $ has no closed form
solutions in terms of simple functions, and then it must be solved by
numerical methods. In this work, we proceed to obtain the the numerical
computation of $\bar{N}$ by a root-finding procedure. By solving this
quantization condition, one obtains 
\begin{equation}
\beta +\frac{\gamma }{\delta }+\frac{1}{2}=\bar{N},  \label{eq19}
\end{equation}%
where $\bar{N}$ now is an undetermined number. The values of the $\beta
,\gamma $ and $\delta $ given by $\left( \ref{eq9b}\right) $ yields the
possible energy levels%
\begin{equation}
\varepsilon _{\pm }=-\omega \left\vert l\right\vert \pm \sqrt{p_{z}^{2}+m^{2}%
\left[ 1-\frac{\eta ^{2}}{\left( -\bar{N}+\frac{1}{2}+\sqrt{l^{2}/\alpha
^{2}+\eta ^{2}}\right) ^{2}}\right] .}  \label{eq20}
\end{equation}%
Similar to the case where $\alpha \omega \ll 1$ , we can see that the
discrete set of energies for both particle ($\varepsilon _{+}$) and
antiparticle ($\varepsilon _{-}$) is composed of two contributions: the
first term is associated to the Coulomb-like potential embedded in a cosmic
string background and the second term is associated to the noninertial
effect of rotating frames, which in turn is a Sagnac-type effect \cite%
{inercial1,inercial2}. From (\ref{eq17}) and (\ref{eq20}), we can see that
energy spectrum is inversely proportional to $\alpha $ for a $\eta $ fixed.
If $\eta $ increases (decreases) for a $\alpha $ fixed, the energy spectrum
decreases (increases). The first numerical values of $\bar{N}$ and their
respective energies are listed in Tables 1 and 2.

\begin{table}[H]
\caption{The first values of $\bar{N}$ \ that satisfy the
quantization condition for $\omega=0.2$.} 
\begin{ruledtabular}
\begin{tabular}{ccccc}

$l$ & $\alpha$ & $\bar{N}$  & $\varepsilon_{+}$ & $\varepsilon_{-}$  \tabularnewline
\hline
0 & 0.9 &  0.000000  & 47.9157  &  -47.9157  \tabularnewline
0 & 0.9 & -1.000000  & 49.6528  &  -49.6528  \tabularnewline
0 & 0.9 & -2.000000  & 49.8630  &  -49.8630  \tabularnewline
0 & 0.9 & -3.000040  & 49.9269  &  -49.9269  \tabularnewline
0 & 0.9 & -4.033890  & 49.9554  &  -49.9554  \tabularnewline
1 & 0.9 &  0.000000  & 49.4217  &  -49.8217  \tabularnewline
1 & 0.9 & -1.000000  & 49.6551  &  -50.0551  \tabularnewline
1 & 0.9 & -2.000020  & 49.7240  &  -50.1240  \tabularnewline
1 & 0.9 & -3.021300  & 49.7537  &  -50.1537  \tabularnewline
1 & 0.9 & -4.774410  & 49.7756  &  -50.1756  \tabularnewline
1 & 0.6 & -0.000000  & 49.5889  &  -50.9889  \tabularnewline
1 & 0.6 & -1.000000  & 49.7009  &  -50.1009  \tabularnewline
1 & 0.6 & -2.000000  & 49.7427  &  -50.1427  \tabularnewline
1 & 0.6 & -3.000771  & 49.7627  &  -50.1627  \tabularnewline
1 & 0.6 & -4.097760  & 49.7746  &  -50.1746  \tabularnewline

\end{tabular}
\end{ruledtabular}
\end{table}

\begin{table}[H]
\caption{The first values of $\bar{N}$ \ that satisfy the
quantization condition for $\omega=0.25$.} 
\begin{ruledtabular}
\begin{tabular}{ccccc}

$l$ & $\alpha$ & $\bar{N}$  & $\varepsilon_{+}$ & $\varepsilon_{-}$  \tabularnewline
\hline
0 & 0.9 &  0.000000  & 47.9157  &  -47.9157  \tabularnewline
0 & 0.9 & -1.000000  & 49.6528  &  -49.6528  \tabularnewline
0 & 0.9 & -2.000000  & 49.8626  &  -49.8626  \tabularnewline
0 & 0.9 & -3.002345  & 49.9270  &  -49.9270  \tabularnewline
0 & 0.9 & -4.287070  & 49.9598  &  -49.9598  \tabularnewline
1 & 0.9 &  0.000000  & 49.3217  &  -49.8717  \tabularnewline
1 & 0.9 & -1.000000  & 49.6051  &  -50.1051  \tabularnewline
1 & 0.9 & -2.001140  & 49.6741  &  -50.1741  \tabularnewline
1 & 0.9 & -3.199530  & 49.7071  &  -50.2071  \tabularnewline
1 & 0.9 & -10.24775  & 49.7429  &  -50.2429  \tabularnewline
1 & 0.6 &  0.000000  & 49.5389  &  -50.0389  \tabularnewline
1 & 0.6 & -1.000000  & 49.6509  &  -50.1509  \tabularnewline
1 & 0.6 & -2.000037  & 49.6927  &  -50.1927  \tabularnewline
1 & 0.6 & -3.025394  & 49.7131  &  -50.2131  \tabularnewline
1 & 0.6 & -4.743860  & 49.7291  &  -50.2291  \tabularnewline

\end{tabular}
\end{ruledtabular}
\end{table}

The first values of $\bar{N}$ \ that satisfy the quantization condition for
arbitrary $\alpha \omega $ are consistent with the results for the limit $%
\alpha \omega \ll 1$. The explanation for this result is found in the
asymptotic behavior of the confluent hypergeometric function $%
_{1}F_{1}\left( A,B;\rho \right) $. If $\rho \rightarrow \infty ,$ the
asymptotic behavior of function is given by%
\begin{equation*}
_{1}F_{1}\left( A,B;\rho \right) \simeq \frac{\Gamma \left( B\right) }{%
\Gamma \left( B-A\right) }e^{-i\pi A}\rho ^{-A}+\frac{\Gamma \left( B\right) 
}{\Gamma \left( A\right) }e^{\rho }\rho ^{A-B}\text{,}
\end{equation*}%
we can see that the presence of $e^{\rho }$ in the second term spoils the
normalizability of $R\left( \rho \right) $ , but if $A=-N$ \ (non-negative
integer) this behavior can be remedied and the normalizability of $R\left(
\rho \right) $ is obtained. The numerical computation of $\bar{N}$ in Table
1 suggests that $\bar{N}$ tends to a non-negative integer as the product $%
\alpha \omega $ is small, i.e, $1/\alpha \omega \rightarrow \infty $. \ 

\section{Conclusions}

In this paper we have determined the Klein-Gordon equation in the presence
of a Coulomb-like scalar potential in a curved spacetime. From these results
a compact expression for the energy spectrum associated with the
Klein-Gordon equation in a cosmic string space has been provided.

We have shown that the noninertial effects restrict the physical region of
the spacetime where the particle can be observed and shift its energy
levels. This feature is an indicator of the coupling between the angular
quantum number and the angular velocity of the rotating frame.

We also have shown that the presence of a Coulomb-like scalar potential
allows the formation of bound states and the energy spectrum associated with
this equation in a cosmic string space depends on the deficit angle of the
conical spacetime.

Due to noninertial effects, the system presents two different classes of
solutions that depend on the value of the product $\alpha\omega$. The first
case is obtained by adopting the limit $\alpha\omega\ll1$, that means a not
so fast rotation, and as a second case, we consider an arbitrary relation $%
\alpha\omega.$ For both classes of solutions, we have found the energy
spectrum and the eigenfunctions.

We have shown that the discrete set of energies is composed of two
contributions. The first term is associated to the Coulomb potential
embedded in a cosmic string background and the second one is associated to
the noninertial effect of rotating frames.

With these results it is possible to have an idea about the general aspects
of the behavior of spin-0 particles inside a cosmic string space where a
scalar potential exists. Potential physical applications of the results of
this paper include the theory of defects in solids involving condensed
matter physics. The idea is to explore the well known analogy between cosmic
strings and disclinations in solids \cite{defects} where the metric which
describes a disclination corresponding to the spatial part of the line
element of the cosmic string.

\bigskip

\section{Acknowledgments}

This work was supported in part by means of funds provided by CAPES.

\bibliographystyle{aipnum4-1}
\bibliography{referencias_tese}

\end{document}